\documentclass[aps, twocolumn,superscriptaddress, pra]{revtex4-1}
\usepackage[utf8]{inputenc}

\usepackage{lipsum}
\usepackage{braket}
\usepackage{amsmath}
\usepackage{amssymb}
\usepackage{hyperref}
\usepackage{siunitx}
\usepackage[normalem]{ulem}
\usepackage{graphicx}
\usepackage{colortbl}
\usepackage{xcolor}

\begin{document}
\title{Dynamical uncertainty propagation with noisy quantum parameters}
\author{Mogens Dalgaard}
\affiliation{Department of Physics and Astronomy, Aarhus University, Ny Munkegade 120, 8000 Arhus C, Denmark}

\author{Carrie A. Weidner}
\affiliation{Department of Physics and Astronomy, Aarhus University, Ny Munkegade 120, 8000 Arhus C, Denmark}

\author{Felix Motzoi}
\email{f.motzoi@fz-juelich.de}
\affiliation{Forschungszentrum J\"ulich, Institute of Quantum Control (PGI-8), D-52425 J\"ulich, Germany}

\date{\today}

\begin{abstract}
   Many quantum technologies rely on high-precision dynamics, which raises the question of how these are influenced by the experimental uncertainties that are always present in real-life settings. A standard approach in the literature to assess this is Monte Carlo sampling, which suffers from two major drawbacks. First, it is computationally expensive. Second, it does not reveal the effect that each individual uncertainty parameter has on the state of the system. In this work, we evade both these drawbacks by incorporating propagation of uncertainty directly into simulations of quantum dynamics, thereby obtaining a method that is faster than Monte Carlo simulations and directly provides information on how each uncertainty parameter influence the system dynamics. Additionally, we compare our method to experimental results obtained using the IBM quantum computers.
\end{abstract}
\maketitle
Technologies based on quantum information may lead to groundbreaking progress within numerical search~\cite{grover1996fast}, cryptography~\cite{shor1999polynomial},  simulation~\cite{barreiro2011open,bloch2012quantum}, optimization~\cite{das2008colloquium, mcgeoch2014adiabatic}, and machine learning~\cite{schuld2015introduction,biamonte2017quantum}. Realization of these technologies requires improvements in our ability to measure, design, build, and realistically model subparts of these quantum systems~\cite{acin2017european}. The latter necessitates the incorporation of the experimental uncertainties that are always present in real quantum systems into numerical simulations of such systems.   

Propagation of uncertainty (or error) is a standard tool for understanding how experimental uncertainties transform in calculations~\cite{taylor1997introduction}. In this work we incorporate propagation of uncertainty into quantum dynamical simulation and characterization by drawing on methods from quantum control theory~\cite{khaneja2005optimal,motzoi2011optimal,de2011second,dalgaard2020hessian}. In doing so we demonstrate a significant speed up relative to Monte Carlo sampling, which is the current standard in the literature (see, e.g., Refs.~\cite{langrock2020reset,sorensen2020optimization,ge2020robust, vieira2020almost,song2021average,estarellas2020comparison,coden2021controlled,cabedo2021excited,kiely2021fast,d2020fast,calderon2021fast,wu2021resilient,giorgi2020topological,hu2020topological} for a series of recent papers). 

Moreover, our method allows for an in-depth analysis of how experimental uncertainties independently or cooperatively influence the dynamical observables in the system. 
This is valuable for investigating the consistency between models and experimental results, as well as the characterization of noise, uncertainty levels, and the influence of different experimental parameters. As one example, within circuit QED we have, in recent years, seen improvements in one- and two-qubit gate operations due to better understanding and calibration of the experimental systems~\cite{sheldon2016procedure,theis2018counteracting,patterson2019calibration,magesan2020effective, sundaresan2020reducing}. Careful study of these systems’ uncertainties and their influence on quantum dynamics is necessary to push this frontier even further. 

The methods proposed in this paper could also help in the construction of composite control sequences that are specifically designed to perform robustly against fluctuations in the system parameters~\cite{wimperis1994broadband,cummins2003tackling,vandersypen2005nmr,mottonen2006high}, as well as enhancing quantum control and optimization protocols~\cite{khaneja2005optimal,borneman2010application, skinner2011design, motzoi2016backaction, platzer2010optimal, sorensen2020optimization, ge2020robust}. That is, the sensitivity to parameter uncertainties can be modified as needed for tomographic reconstructions \cite{d2001quantum,dankert2009exact}, system identification \cite{mabuchi1996dynamical, schirmer2010quantum, burgarth2012quantum, zhang2014quantum, guctua2015system}, and quantum metrology \cite{braunstein1994statistical,giovannetti2004quantum,giovannetti2011advances}. In addition, the presented method could also aid physicists in designing quantum computation architectures~\cite{koch2007charge,rigetti2010fully,goerz2017charting,brooks2013protected,menke2021automated,nguyen2019high}. These need to perform stably under smaller fluctuations in the system parameters, and as such, a detailed analysis of their performance under realistic and expected uncertainties could help assess the robustness of a given architecture. 
 We term our approach Quantum Uncertainty Propagation In Dynamics (QUPID).
 
 This paper is organized as follows: in Section~\ref{sec:theory} we introduce QUPID, in Section~\ref{sec:MC} we compare to Monte Carlo simulations, and in Section~\ref{sec:expm} we compare to experimental results. Further, in Section~\ref{sec:scalling} we investigate the scaling of QUPID and finally, in Section~\ref{sec:outlook}, we conclude the paper. 

\begin{figure*}
    \centering
    \includegraphics{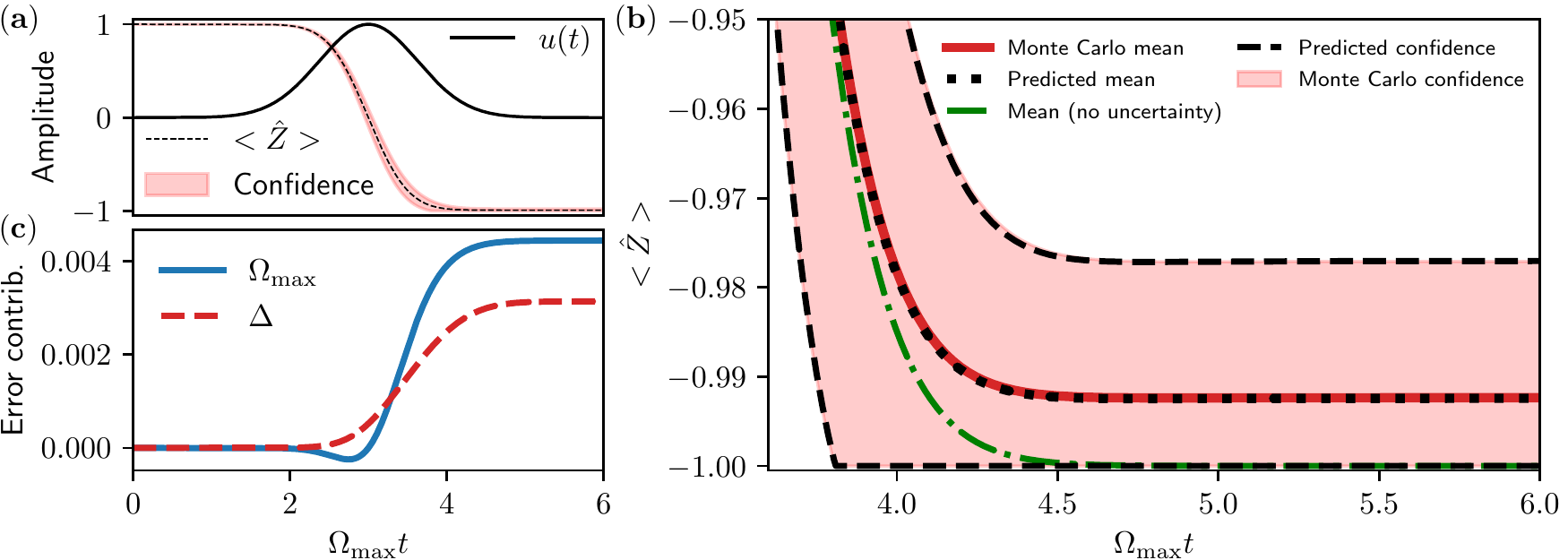}
    \caption{Quantum uncertainty propagation when applied to a Gaussian pulse. (a) The uncertain Gaussian pulse (solid, see main text for details) and the result of Monte Carlo sampling (dashed line, with $2\sigma$ confidence interval shaded in red). (b) Comparison between propagation of uncertainty and Monte Carlo sampling, zoomed in on later times. Here, the Monte Carlo mean (solid line) and confidence interval (shading) are shown in red, and the mean and confidence intervals derived from our method are shown with the black dotted and dashed lines, respectively, showing excellent agreement relative to the Monte Carlo case. The mean for the uncertainty-free case is shown by the green dash-dotted line (c) The relative error contribution as a function of time, i.e., the contribution to the shift from the uncertainty-less case resulting from uncertainties in the maximum drive amplitude $\Omega_\mathrm{max}$ (blue, solid) and detuning $\Delta$ (red, dashed).}
    \label{fig:gaussian}
\end{figure*}

\section{Theory} \label{sec:theory}
Let $Z(t)$ denote a function that depends on the evolution of the system, such as the expectation value of a Hermitian operator $Z(t) = \braket{\hat Z}$, or the Haar-averaged gate fidelity $Z(t) = \int_\psi\bra{\psi}U(t)^\dagger U_{target}\ket{\psi}d\psi $. Now let the Hamiltonian operator and other system evolution variables depend on a set of normally distributed parameters $\boldsymbol{\theta} = (\theta_1, \theta_2, \ldots, \theta_M)$ with mean $\bar{\boldsymbol{\theta}} = (\bar{\theta_1}, \bar{\theta_2}, \ldots,\bar{\theta_M})$ and covariances $\sigma_{i,j} = \mathbb{E}[(\theta_i - \bar{\theta_i})(\theta_j - \bar{\theta_j})]$, which we assume are relatively small~\footnote{Note that the expression we derive for the mean works for any symmetric distribution.}. Here $\sigma_j^2 = \sigma_{j,j}$ denotes the variance. Thus, $Z$ is a function of the parameters $\boldsymbol{\theta}$, which we may Taylor expand around the mean values $\bar{\boldsymbol{\theta}}$,
\begin{align}\nonumber
    Z(\boldsymbol{\theta}, t) &\simeq Z(\bar{\boldsymbol{\theta}}, t) + \sum_j \frac{\partial Z(\bar{\boldsymbol{\theta}},t)}{\partial \theta_j} (\theta_j - \bar{\theta}_j)
    \\
    &+ \frac{1}{2} \sum_{i,j} \frac{\partial^2 Z(\bar{\boldsymbol{\theta}}, t)}{\partial \theta_i \partial \theta_j} (\theta_i - \bar{\theta}_i) (\theta_j - \bar{\theta}_j),
    \label{eq:second_order_expansion}
\end{align}
where truncating the Taylor expansion to second order is justified by assuming small covariances. Evaluating the mean value reveals

\begin{align}
    \mathbb{E}[Z(\boldsymbol{\theta},t)] 
    &\simeq Z(\boldsymbol{\bar{\theta}},t) 
    + \frac{1}{2} \sum_{i,j} \frac{\partial^2 Z(\bar{\boldsymbol{\theta}},t)}{\partial \theta_i \partial \theta_j} \sigma_{i,j}.
    \label{eq:mean_value}
\end{align}
From Eq.~(\ref{eq:mean_value}) we can see directly how uncertainties in the system dynamics may lead to shifts away from the expected value $Z(\boldsymbol{\bar{\theta}},t)$ at different times. For example, this term could cause a reduction in the predicted average gate fidelity. This is especially relevant to analytical protocols for parameter estimation, where stochasticity in the parameter may itself modify the mean estimation. Moreover, the intuitive approach of assessing robustness via first derivatives will fail to capture the expected loss in average fidelity, as its effect will simply average to zero. In the case that the uncertainties are statistically independent ($\sigma_{i,j} = 0$ if $i\neq j$), we can further see that the individual parameter uncertainties are additive, allowing them to be calculated individually by, e.g., Monte Carlo simulations, or more efficiently, as we show below, by propagation of uncertainty through the dynamics. 

A similar calculation for the variance (see Supplemental Material) yields 
\begin{align} \nonumber
\text{Var}[Z(\boldsymbol{\theta}, t)]
&\simeq
\sum_{i,j} 
\frac{\partial Z(\bar{\boldsymbol{\theta}},t)}{\partial \theta_i}
\frac{\partial Z(\bar{\boldsymbol{\theta}},t)}{\partial \theta_j} 
\sigma_{i,j} \\
&+
\frac{1}{2}
\sum_{i,j,k,l}
\frac{\partial^2 Z(\bar{\boldsymbol{\theta}},t) }{\partial \theta_i \partial \theta_j}
\frac{\partial^2 Z(\bar{\boldsymbol{\theta}},t) }{\partial \theta_k \partial \theta_l}
\sigma_{i,k} \sigma_{j,l}.
\label{eq:variance}
\end{align}
We now sketch how to obtain the derivatives that appear in Eq.~(\ref{eq:mean_value}) and (\ref{eq:variance}) through analysis of the dynamics. 

In order to be able to calculate the dynamical uncertainties in system observables with the above equations, it is necessary to be able to efficiently propagate the equations of motion for the derivatives of these observables.  To do this, we draw on established methods from the field of quantum optimal control theory~\cite{khaneja2005optimal,de2011second,motzoi2011optimal,dalgaard2020hessian}. We have so far tacitly assumed that $Z$ was a functional of the current state of the system $\chi(t)$, which solves the dynamic equation of motion $\frac{\partial }{\partial t} \chi(t) = A(t) \chi(t)$. This could be, e.g., the Schrödinger equation $A(t) = -i H(t)$ with $\hbar = 1$ and $\chi$ being a single quantum state, a unitary operator, or a collection of states, or a density matrix with evolution governed by the von Neumann or the Lindblad master equation. We elaborate further on this in the Supplemental Material. 

We discretize the time evolution for a total duration $T$ in $N$ equidistant steps $\Delta t$ such that $\Delta t = T/N$. This allows us to numerically solve the equation of motion using the time evolution operator $\chi_{n+1} = U_n \chi_n$, where $\chi(t_n) = \chi_n$ and $U_n = e^{A(t_n + \Delta t/2)}$ is the mid-point interpolation of the truncated Magnus series~\cite{blanes2009magnus}, leading to a global second order-error in $\chi(T)$ that can be made arbitrarily small by suitable choice of $\Delta t$. The derivatives of $Z$ ultimately depend on the derivative of $\chi$, which we may obtain by using the chain rule

\begin{align}
    \frac{\partial}{\partial \theta_j} \chi_{n+1} 
    = 
    \Big( \frac{\partial}{\partial \theta_j} U_n \Big) \chi_n + U_n \frac{\partial}{\partial \theta_j} \chi_n,
\end{align}
and similarly we obtain
\begin{align} \nonumber
    \frac{\partial^2}{\partial \theta_i \partial \theta_j} \chi_{n+1} &= 
    \Big( \frac{\partial^2}{\partial \theta_i \partial \theta_j} U_n \Big) \chi_n
    +
    \Big( \frac{\partial}{ \partial \theta_j} U_n \Big) \frac{\partial}{ \partial \theta_i} \chi_n\\
    &+ \Big(\frac{\partial}{\partial \theta_i} U_n \Big) \frac{\partial}{\partial \theta_j} \chi_n
    + U_n \frac{\partial^2}{\partial \theta_i \partial \theta_j} \chi_n.
\end{align}

To calculate the derivatives for each time slice one should solve~\cite{wilcox1967exponential} 
\begin{equation}
     \frac{d}{d \eta} e^{\chi(\eta)} = \int_0^{1} e^{\alpha \chi(\eta)} \frac{d \chi(\eta)}{d \eta} e^{(1-\alpha) \chi(\eta)} d \alpha.
    \label{eq:matrix_derivative}
\end{equation}
which has a known solution in terms of the eigenvalues of $\chi$~\cite{de2011second,machnes2011comparing,dalgaard2020hessian}. Further steps in the derivation are given in the Supplemental Material.

\begin{figure*}
    \centering
    \includegraphics{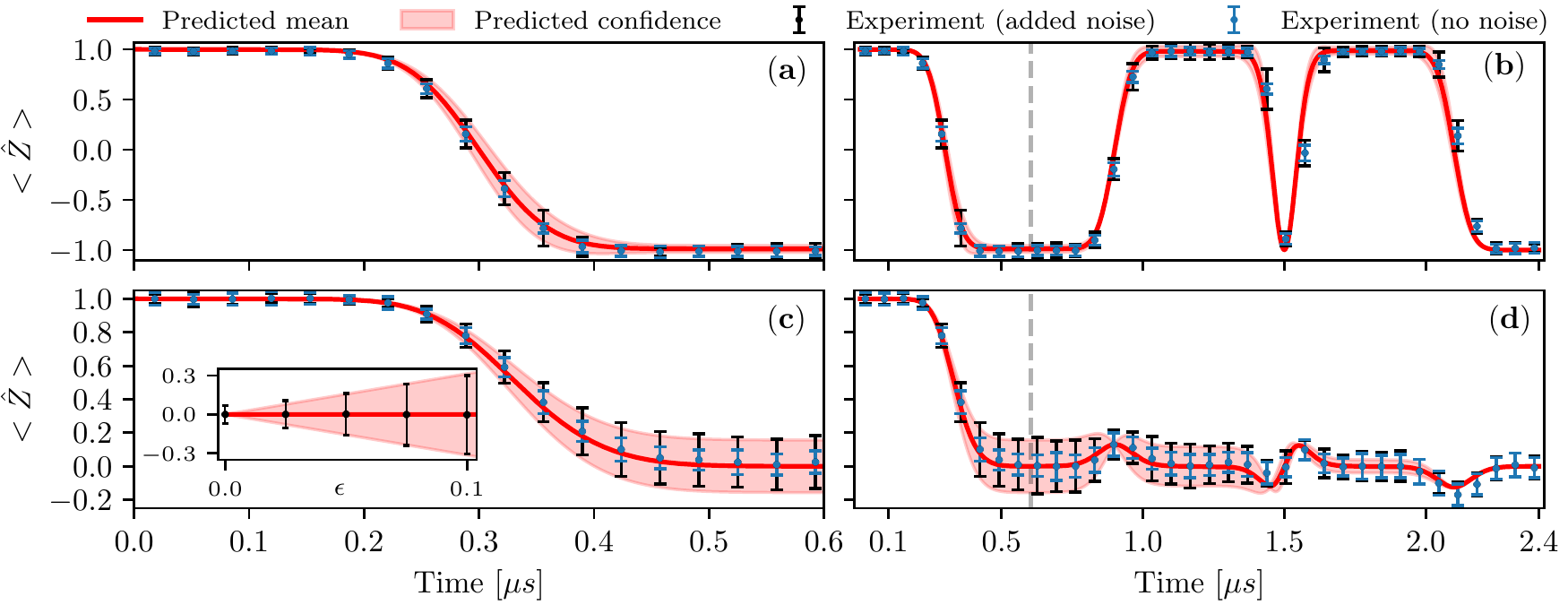}
    \caption{Comparison of QUPID to experimental results using the IBM Armonk qubit showing how the uncertainty changes over time for a Gaussian (a) and BB1 (b) $\pi$ pulse and a Gaussian (c) and BB1 (d) $\pi/2$ pulse. The black points (error bars) denote the mean (twice the standard deviation of   the mean) for $50$ experimental runs, where for each run, we sample the amplitude from a normal distribution of width $\epsilon = 0.05$. The blue points and error bars were generated from data taken with a constant amplitude, i.e., $\epsilon = 0$. In both cases, each experimental run outputs the average results over $1024$ measurements. The theoretical BB1 uncertainty goes to zero at the endpoints since the BB1 pulses are robust up to sixth-order error in amplitude, but otherwise we obtain excellent agreement between theory and experiment when the amplitude uncertainty errors dominate over other errors in the system, e.g., measurement errors. The inset in (c) compares the predictions of the final expectation value $\braket{Z}$ to experimental data with respect to changes in the amplitude error $\epsilon$ for a Gaussian $\pi/2$ pulse. Plots (a) and (c) zoom in on the first Gaussian pulse of the BB1 sequence (marked with a vertical dashed line); some data points have been omitted in the BB1 plots for clarity.} 
    \label{fig:gaussian_expm}
\end{figure*}

\section{Comparison to Monte Carlo simulations} \label{sec:MC}
We consider here a two-level Hamiltonian given by ($\hbar = 1$) 
\begin{align} 
    H(t) = \Delta \hat{Z} + \Omega_{\max} u(t) \hat{X}, 
    \label{eq:H} 
\end{align} 
with detuning  $\Delta = 0$, maximum amplitude $\Omega_{\max}$, Pauli operators $\hat{X}$ and $\hat{Z}$, and dimensionless control function $-1\leq u(t) \leq +1$. We consider a state transfer $\ket{0} \rightarrow \ket{1}$ where the dynamical evolution is subject to uncertainties in the system parameters $\Omega_{\max}$ and $\Delta$. The transfer itself is induced by Gaussian $\pi$-pulse (Fig.~\ref{fig:gaussian}(a)). We time-evolve the system for a duration $T = 6/\Omega_{\max}$ with independent uncertainties $\sigma(\Omega_{\max}) / \Omega_{\max} = \sigma(\Delta) / \Omega_{\max} = 3.0 \times 10^{-2}$ using 10,000 Monte Carlo simulations. We calculate the average trajectory over the Pauli observable $\braket{\hat{Z}}$ and define the confidence interval as twice the standard deviation $\pm 2\sigma(\braket{\hat{Z}})$ from the average trajectory, which we truncate between $\pm 1$, as this is a physical limitation on $\braket{\hat{Z}}$ (Fig.~\ref{fig:gaussian}(a)). We now compare the predictions of QUPID, Eqs.~(\ref{eq:mean_value}) and (\ref{eq:variance}), to the Monte Carlo simulations by zooming in on the end of the trajectory in Fig.~\ref{fig:gaussian}(b). Here we see how uncertainties in the system parameters induce a shift in the mean value away from the ideal, uncertainty-less case $\braket{\hat{Z}} = -1$. In addition, we see an excellent agreement between the Monte Carlo simulations and uncertainty propagation predictions. On a standard laptop computer, the Monte Carlo simulations take $\approx 30$ minutes, whereas QUPID takes $\approx 0.2$ seconds, thereby providing a significant speed up. 

An additional benefit of QUPID over Monte Carlo methods is that Eq.~(\ref{eq:mean_value}) allows us to directly track how each uncertainty parameter influences the system dynamics. To this end, we may define the error contribution from Eq.~(\ref{eq:mean_value}) for each parameter as $\frac{1}{2} \frac{\partial^2 \braket{\hat{Z}}}{\partial \theta^2}  \sigma(\theta)^2$ with $\theta =\Omega_{\max}, \Delta$. We plot in Fig.~\ref{fig:gaussian}(c) the error contribution for each parameter along the trajectory. Interestingly, we see that around half-way through the trajectory, the error contributions from the amplitude and detuning are opposite in sign and thus cancel each other out to some extent; such fortuitous cancellations can be analyzed and exploited when designing pulse sequences using our method. We also see at the end of the trajectory that the two error contributions converge on different specific values, i.e.~uncertainty in the amplitude and detuning constitute respectively around $59\%$ and $41\%$ of the total error. This analysis is useful in determining, for example, which parameters are most sensitive in an experiment and thus where efforts must be concentrated when improving the precision of experimental systems.

\section{Experimental comparison} \label{sec:expm}
We compare the predictions of QUPID to a single-qubit experiment via pulse-level control of IBM's \emph{ibmq\textunderscore armonk} single-qubit system accessible via Qiskit pulse~\cite{QiskitOpenPulse, QiskitOpenPulse2, McKay_2020}. Such pulse-level control has previously been used to optimize the fidelity of cross-resonance gates~\cite{sheldon2016procedure}, the CV-gate~\cite{Yamamoto_2021}, and to build a quantum compiler implementing basis gates and qudit operations~\cite{Chong_2020}.

First, we determine Gaussian and composite BB1~\cite{wimperis1994broadband} pulse sequences that drive  $\pi/2$ (Hadamard) and $\pi$ (bit-flip) transitions. After calibration (see Supplemental Material), we artificially add amplitude noise to each of the aforementioned pulse sequences. The amplitude noise is sampled from a Gaussian distribution centered around the optimal amplitudes with a width of $\epsilon = \sigma(\Omega_\mathrm{max})/\Omega_\mathrm{max}$ for $\Theta = \pi/2$ and $\pi$~\footnote{For the qubit drive times considered here, the intrinsic qubit errors, e.g. due to relaxation, are negligible, so this system as designed is a suitable playground for artificially-inserted uncertainties that can be used to test our theory.}. We compare the predictions of our theory to the experimental results in Fig.~\ref{fig:gaussian_expm} with $\epsilon = 0.05$ for Gaussian and BB1 $\pi$ and $\pi/2$ pulses. The data and simulation match well throughout the duration of the pulses for both cases, although our theory cannot fully account for measured BB1 errors at the final time due to discrimination/shot noise (see blue error bars) being larger than the BB1 final error, which has intrinsic robustness up to sixth order in amplitude error~\cite{wimperis1994broadband, vandersypen2005nmr}. To determine how the error scales with $\epsilon$, we compare the experimental data and theoretical predictions for the final expectation value $\braket{\hat{Z}}$ of a Gaussian $\pi/2$ pulse in the inset of Fig.~\ref{fig:gaussian_expm}(c), where we again obtain excellent agreement between theory and experiment. Thus, when compared with experimental data, our theory can determine at what point various experimental uncertainties dominate the error, which can be useful for experimental debugging, theoretical model-building, or designing new systems. Moreover, by calibrating the exact value of $\epsilon$ in the model, one can readily use Eq.~\eqref{eq:mean_value} to remove the correction term coming from uncertainty and thereby estimate the true model parameters, e.g. for Hamiltonian learning or quantum sensing.

\begin{figure}
    \centering
    \includegraphics{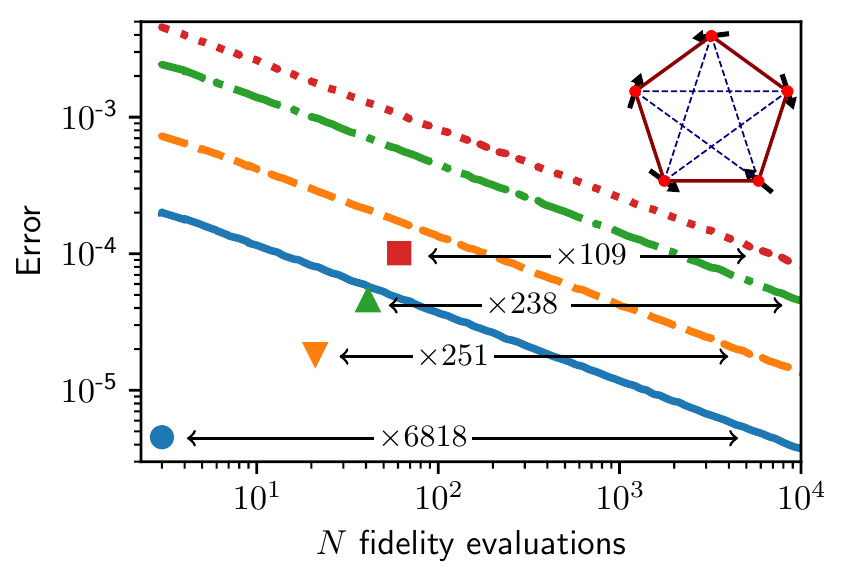}
    \caption{Simulation results with the spin-star system (top right inset). The lines depict the expected error of Monte Carlo simulations as a function of the number of fidelity evaluations for 1 (blue solid), 10 (orange dashed), 20 (green, dash-dotted), and 30 (red dotted) parameters. This is compared to QUPID for 1 (blue circle), 10 (orange, down triangle), 20 (green, up triangle) and 30 (red square) parameter(s) chosen at random, where the computational expense has been evaluated with respect to a finite difference scheme. The horizontal arrows mark the speed improvement in an equal-accuracy comparison.}
    \label{fig:spin_star}
\end{figure}

\section{Scaling and speed up} \label{sec:scalling}
Here, we consider how QUPID scales with the number of parameters compared to Monte Carlo simulations. In the case of independent uncertainties, estimating the mean via Eq.~(\ref{eq:mean_value}) scales linearly with the number of parameters, while it scales quadratically in the more general case of dependent uncertainties. If we compare to Monte Carlo simulations, the averaging error over $N$ samples is given by the standard deviation of the mean $\bar{ \sigma} = \sigma/\sqrt{N}$, leading to a $N^{-1/2}$ scaling. 
We may consider the scaling of the standard deviation by considering the square root of Eq.~(\ref{eq:variance}). In the general case of dependent uncertainties, Eq.~(\ref{eq:variance}) has a fourth order scaling with the number of parameters, which implies that the standard deviation would scale quadratically. In the case of independent parameters, this reduces to a linear scaling. In other words, the scaling of the two approaches with respect to the number of uncertainty parameters remains, in principle, the same. 

However, in practice, many of the mixed derivatives in Eq.~(\ref{eq:variance}) tend to be zero, leading to an effective sub-linear and sub-quadratic scaling of Monte Carlo simulations with independent and dependent uncertainties, respectively. For this reason, we should compare the precision of the two methods over a wide range of uncertainty parameters. We consider the spin-star system illustrated in the top right of Fig.~\ref{fig:spin_star}. The system consists of 5 qubits with interactions between the nearest and the next-nearest spins, where we model an external, global $X$-control for all qubits. The Hamiltonian for the spin-star system is given by 
\begin{align} \nonumber
    H &= u(t) \sum_j \hat{X}_j \\ \nonumber
    &- \sum_{\braket{i,j}} 
    \big(
    J_x^{(i,j)} \hat{X}_i \hat{X}_j
    + J_y^{(i,j)} \hat{Y}_i \hat{Y}_j
    + J_x^{(i,j)} \hat{Z}_i \hat{Z}_j
    \big) \\
    & -\sum_{\langle \langle i,j \rangle \rangle}
    \big( g_x^{(i,j)} \hat{X}_i \hat{X}_j
    + g_y^{(i,j)} \hat{Y}_i \hat{Y}_j
    + g_x^{(i,j)} \hat{Z}_i \hat{Z}_j \big),
\end{align}
where $\braket{,}$ denotes the sum over neighboring qubits, while $\braket{ \braket{,} }$ denotes next-neighboring qubits. $J = J_k^{(i,j)}$ denotes the nearest coupling and we choose the next-nearest coupling $g = g_k^{(i,j)} = J/10$ for $k = x,y,z$ and all indices $(i,j)$. With this model we have $30$ different parameters. We start in the initial state $\ket{\psi_0} = \ket{0,0,0,0,0}$ and evolve with a control pulse of the form $u(t) = \phi_1(t) + \phi_3(t) - \phi_5(t)$, where $\phi_n(t) = \sin(n\pi t /T)$ for a total duration $TJ = 10$. In the absence of uncertainties this leads to a final state $\ket{\tilde{\psi}} = \ket{\psi(T)}$ that populates a wide range of different Fock states in the Hilbert space. 

We investigate how uncertainties in the system parameters induces population in states other than $\ket{\tilde{\psi}}$, which can be described by the fidelity $F = \braket{P}$, with $P = \ket{\tilde{\psi}} \bra{\tilde{\psi}}$ denoting the projector onto the expected final state. Again, we assume independent normally-distributed uncertainties, with relative amplitude $\sigma(J)/J = \sigma(g)/g = 0.01$, which we model in a random subset of the $30$ system parameters. This generally leads to a population leakage away from the uncertainty-less case with an error $1-F$ in the range from $10^{-2} $ to $10^{-4}$ depending on the number of parameters. 

We model uncertainties in $1$, $10$, $20$, and $30$ parameters chosen at random, where we perform 100,000 Monte Carlo simulations for each choice. To understand the scaling of the Monte Carlo method, we compare the difference between the fidelity average over the 100,000 samples ($F_{\text{true}}$) versus only averaging over a subset of the simulations ranging from $1$ to 10,000 ($F_{\text{approx}}$).  

For each given subset, we draw a random number of fidelities from the entire set of simulated fidelities. This procedure is repeated 10,000 times in order to calculate the mean estimation error i.e., average over $|F_{\text{true}}-F_{\text{approx}}|$. The results are depicted in Fig.~\ref{fig:spin_star} for different numbers of uncertainty parameters. Note the lines scale as $N^{-1/2}$, consistent with the standard deviation of the mean, but at different heights due to different standard deviations over the fidelity distributions. 

We may compare the performance of QUPID to Monte Carlo simulations in both accuracy and computational expense. The latter is achieved by comparing the number of fidelity evaluations required to calculate the derivatives using a finite difference scheme (see Supplemental Material). Since evaluating the fidelities is by far the most expensive step in each approach, this gives an accurate comparison. We plot the results of QUPID in Fig.~\ref{fig:spin_star} as individual data points.   

We may read the speed up of QUPID relative to Monte Carlo simulations in an equal-accuracy comparison by considering the distance from the individual data points in Fig.~\ref{fig:spin_star} to the lines along the first axis. For clarity we have indicated the factor of speed up with arrows on the figure, which ranges from $6818$ for one parameter to $109$ for $30$ parameters. Note that the relative speed up grows smaller with the number of parameters due to the sub-linear scaling of Monte Carlo. This implies that for a large enough number of uncertainty parameters, Monte Carlo would be preferable. However, we have still demonstrated significant improvements with up to 30 parameters, which would comprise the vast majority of cases typically considered in the literature~\cite{langrock2020reset, vieira2020almost,song2021average,estarellas2020comparison,coden2021controlled,cabedo2021excited,kiely2021fast,d2020fast,calderon2021fast,wu2021resilient,giorgi2020topological,hu2020topological}.  

\section{Outlook} \label{sec:outlook}
In this work, we have demonstrated QUPID, a method for analytically incorporating and propagating uncertainties during quantum dynamical evolution. For a reasonable number of parameters (at least up to 30), this method performs dramatically better than the standard, cutting-edge methods that are commonly used in the literature, namely Monte Carlo sampling. This analytical form has the additional advantages of (a) being able to separate the role of different error sources and (b) understanding how these uncertainties develop over time. This can be useful for theoretical protocol improvement but also for experimental design and debugging. That is, with an understanding of the dominant uncertainty sources in a given protocol, experimentalists can better understand the level of precision that must be reached in their system and which parameters are most likely to cause imprecisions. Likewise, these analytical forms allow the design of protocols (e.g. pulse sequences) that may either decrease or increase the effect of parameter uncertainty, which is useful for increased robustness or improved sensing, respectively.

\acknowledgements
We acknowledge the use of IBM Quantum services for this work. The views expressed are those of the authors, and do not reflect the official policy or position of IBM or the IBM Quantum team. This work was funded by the Carlsberg Foundation and by the Deutsche Forschungsgemeinschaft (DFG, German Research Foundation) under Germany's Excellence Strategy – Cluster of Excellence Matter and Light for Quantum Computing (ML4Q) EXC 2004/1 – 390534769. The numerical results presented in this work were obtained at the Centre for Scientific Computing, Aarhus, phys.au.dk/forskning/cscaa.

\onecolumngrid

\section*{Supplementary Material}
\appendix

\subsection*{Calculation of variance} \label{sec:variance}

We start by evaluating the general expression of the variance based on the results derived in the main text
\begin{align}\nonumber
\text{Var}[Z(\boldsymbol{\theta})] &= \mathbb{E} \Big[ \big( Z(\boldsymbol{\theta},t) -  \mathbb{E}[Z(\boldsymbol{\theta}, t)]  \big)^2 \Big]
\\
&= \nonumber
\mathbb{E} \Big[
\sum_{i,j} 
\frac{\partial Z(\boldsymbol{\bar{\boldsymbol{\theta}}}, t)}{\partial \theta_i}
\frac{\partial Z( \boldsymbol{\bar{\boldsymbol{\theta}}}, t)}{\partial \theta_j} 
(\theta_i - \mu_i)
(\theta_j - \mu_j)\\ \nonumber
&+
\sum_{i,j,k} 
\frac{\partial Z( \boldsymbol{\bar{\boldsymbol{\theta}}}, t)}{\partial \theta_k}
\frac{\partial^2 Z \boldsymbol{\bar{\boldsymbol{\theta}}}, t)}{\partial \theta_i \partial \theta_j}
(\theta_k - \mu_k)
(\theta_i - \mu_i)
(\theta_j - \mu_j)\\ \nonumber
&-
\sum_{i,j,k}
\frac{\partial Z( \boldsymbol{\bar{\boldsymbol{\theta}}}, t)}{\partial \theta_k}
\frac{\partial^2 Z( \boldsymbol{\bar{\boldsymbol{\theta}}}, t)}{\partial \theta_i \partial \theta_j}
\sigma_{i,j}
(\theta_k - \mu_k)\\ \nonumber
&+
\frac{1}{4}
\sum_{i,j,k,l}
\frac{\partial^2 Z( \boldsymbol{\bar{\boldsymbol{\theta}}} , t)}{\partial \theta_i \partial \theta_j}
\frac{\partial^2 Z( \boldsymbol{\bar{\boldsymbol{\theta}}}, t)}{\partial \theta_k \partial \theta_l}
(\theta_i - \mu_i)
(\theta_j - \mu_j)
(\theta_k - \mu_k)
(\theta_l - \mu_l)
\\
&-
\frac{1}{2}
\sum_{i,j,k,l}
\frac{\partial^2 Z(\boldsymbol{\bar{\boldsymbol{\theta}}}, t)}{\partial \theta_i \partial \theta_j}
\frac{\partial^2 Z(\boldsymbol{\bar{\boldsymbol{\theta}}}, t)}{\partial \theta_k \partial \theta_l}
\sigma_{k,l}
(\theta_i - \mu_i)
(\theta_j - \mu_j)
+
\frac{1}{4}
\sum_{i,j,k,l}
\frac{\partial^2 Z(\boldsymbol{\bar{\boldsymbol{\theta}}},t)}{\partial \theta_i \partial \theta_j}
\frac{\partial^2 Z(\boldsymbol{\bar{\boldsymbol{\theta}}},t)}{\partial \theta_k \partial \theta_l}
\sigma_{i,j}
\sigma_{k,l}
\Big].
\label{eq:appendix_variance}
\end{align}
In order to evaluate the above we use Wick's probability theorem \cite{wick1950evaluation} (also known as Isserlis' theorem \cite{isserlis1918formula}), which states that the mean of product of $n$ zero-mean normal distributed variables $\{Y_j\}_j^n$ is given by
\begin{align}
\mathbb{E} \Big[Y_1 Y_2 \ldots Y_n\Big]
= \sum_{p\in P_n^2} \prod_{(i,j)\in p}
\sigma_{i,j}
.
\end{align}
Here the sum is over all distinct partitions $P_n^2$ of $\{Y_j\}_j^n$ into pairs. Since this cannot be done for an odd number $n$, the above gives zero for all odd $n$. From this we e.g. have

\begin{align}
    \mathbb{E} \Big[Y_1 Y_2 Y_3\Big] = 0; \quad
    \mathbb{E} \Big[Y_1 Y_2 Y_3 Y_4\Big] = 
    \sigma_{1,2}\sigma_{3,4}
    +\sigma_{1,3}\sigma_{2,4}
    +\sigma_{1,4}\sigma_{2,3}.
\end{align}
Applying Wick's probability theorem to Eq.~(\ref{eq:appendix_variance}) and using that the order of differentiation does not matter we obtain the experssion for the variance as stated in the main text.

\subsection*{Numerical calculation of derivatives} \label{sec:derivatives}

In this work we seek to solve the equation of motion for a quantum system on the form 
\begin{align}
    \frac{\partial}{\partial t} \chi(t) = A(t) \chi(t),
    \label{eq:quantum_time_equation}
\end{align}
which could be the Schrödinger equation with $A(t) = -i H(t)$ and $\chi(t)$ being a single state $\chi(t) = \ket{\psi(t)}$, a unitary $\chi(t) = U(t)$, or a collection of $K$ states with the $n$th column of $\chi$ being the $n$th state  $\chi(t) = \big[\ket{\psi_1(t)}, \ket{\psi_2(t)}, \ldots, \ket{\psi_K(t)} \big]$. Eq.~(\ref{eq:quantum_time_equation}) could also be the von Neumann equation or the master equation on the Linblad form, where $A(t) = \mathbb{L}(t)$ denotes the Liouvillian and $\chi = \ket{\rho}$ the vectorized form of the density matrix i.e. $\ket{\rho}_{n+(m-1)d} = \rho_{n,m}$, with $d$ denoting the Hilbert space dimension \cite{navarrete2015open}. In either case we may numerically solve the equation of motion by propagating the state in sufficiently small time-steps $\Delta t$ via the time evolution operator elaborated on in the main text.


In order to evaluate the derivatives, we assume the time propagator on the form $A(t) = A_0(t) + \sum_n u_n A_n(t)$, where $u_n$ is assumed a constant with time. This could either be a system-parameter $u_n = \theta_n$ or a function of the system parameters. In the latter case we may evaluate the derivative via the chain rule $\frac{\partial}{\partial \theta_j} = \frac{\partial u_n}{\partial \theta_n} \frac{\partial}{\partial u_n}$. Here both the first \cite{de2011second,machnes2011comparing} and second order derivatives \cite{dalgaard2020hessian} has previously been calculated in literature but the results are reproduced here for completeness. Note this approach also allow us model uncertainties in the initial state of the system $\chi_0 = \chi(t_0)$, which could e.g. stem from initialization errors. Note the following works for Hermitian dynamics such as the Schrödinger equation $A = -iH$. For non-Hermitian dynamics the second order derivative is calculable by other means \cite{goodwin2016modified}. In the following we assume, for simplicity, that $A = -iH$.  

The derivative of the time evolution operator may be evaluated in the eigenbasis of $H_n = H(t_n + \Delta t/2)$ \cite{de2011second,machnes2011comparing}

\begin{align}
    \Braket{h|\frac{\partial U_n}{\partial \theta_j}|k}
    = 
    \Braket{h|\frac{\partial H_n}{\partial \theta_j}|k} I(h,k).
\end{align}
Here $\ket{h}$ and $\ket{k}$ denote eigenstates of $H_n$. We have further defined

\begin{align}
    I(h,k) = 
    \begin{cases}
    -i \Delta t e^{-i E_h \Delta t}, &\text{ if } E_h = E_k\\
    \frac{e^{-i E_h \Delta t} - e^{-i E_k \Delta t}}{E_h - E_k}
    , &\text{ if } E_h \neq E_k.
    \end{cases}
\end{align}
Here $E_h$ denotes the eigenenergy of $\ket{h}$. In Ref. \cite{dalgaard2020hessian} the second order derivative was also calculated

\begin{align} \nonumber
    &\Braket{h|\frac{\partial U_n}{\partial \theta_i \partial \theta_j}|k}
    =
    \\
    &\sum_l
    \Big(
    \frac{\partial H_n^{(h,l)}}{\partial \theta_i}
    \frac{\partial H_n^{(l,k)}}{\partial \theta_j}
    +
    \frac{\partial H_n^{(h,l)}}{\partial \theta_j}
    \frac{\partial H_n^{(l,k)}}{\partial \theta_i}
    \Big)
    \mathcal{I}(k,l,h),
\end{align}
where we use the notation $\frac{\partial H_n^{(h,l)}}{\partial \theta_i} = \braket{h| \frac{\partial H_n}{\partial \theta_i} |l}$. Here $\mathcal{I}(k,l,h)$ can be expressed through the already calculated elements $I(h,k)$, see Eq. (B9) in Ref. \cite{dalgaard2020hessian}.

The method outlined here requires propagating a single state, if we assume $M$ system-parameters, we also need $M$ first order derivatives, and $M^2$ second order derivatives. However, since the order of differentiation does not matter, this reduces to $\frac{1}{2}M(M-1)$ second order derivatives. In most applications, this is significantly lower than Monte Carlo sampling, which typically requires several hundreds if not thousands of samples. 

To concretize these ideas, we demonstrate how to calculate the derivatives of an expectation value of a Hermetian operator $\hat{A}$ with respect to a single state $Z(t) = \braket{\psi(t)|\hat{A}|\psi(t)}$ as also treated in the main text. This has also previously been calculated in e.g. Ref. \cite{khaneja2005optimal}.  

\begin{align}\nonumber
    \frac{\partial}{\partial \theta_j} Z(t)
    &= 
    \Braket{\frac{\partial}{\partial \theta_j} \psi(t)|\hat{A}|\psi(t)}
    +
    \Braket{\psi(t)|\hat{A}|\frac{\partial}{\partial \theta_j} \psi(t)}
    +
    \Braket{\psi(t)|\frac{\partial \hat{A}}{\partial \theta_j}|\psi(t)}\\
    &= 2 \mathbb{R}
    \Bigg[
    \Braket{\psi(t)|\hat{A}|\frac{\partial}{\partial \theta_j} \psi(t)}
    \Bigg]
    +
    \Braket{\psi(t)|\frac{\partial \hat{A}}{\partial \theta_j}|\psi(t)}.
\end{align}
Where we in the above equation, have taken into account that $\hat{A}$ itself may also be dependent on the parameters, for instance, the case where $\hat{A}=H$ is the expected energy of the system. The second derivative reads
\begin{align} \nonumber
    \frac{\partial^2}{\partial \theta_i \partial \theta_j} Z(t)
    &=
    2 \mathbb{R}
    \Bigg[
    \Braket{\frac{\partial}{\partial \theta_i} \psi(t)|\hat{A}|\frac{\partial}{\partial \theta_j} \psi(t)}
    +
    \Braket{\psi(t)|\hat{A}| \frac{\partial^2}{\partial \theta_i \partial \theta_j} \psi(t)}
    +
    \Braket{\psi(t)|\frac{\partial \hat{A}}{\partial \theta_j}|\frac{\partial }{\partial \theta_i}\psi(t)}
    \Bigg]\\
    &+ \Braket{\psi(t)|\frac{\partial^2 \hat{A}}{\partial \theta_i \partial \theta_j} |\psi(t)}.
\end{align}

As an alternative to calculating the derivatives exactly, we also demonstrate how they may be estimated using finite differences \cite{fornberg1988generation}. The first derivative using a central difference scheme is given by

\begin{align}
    \frac{\partial Z(\theta_j)}{\partial \theta_j}
    = \frac{Z(\theta_j + \epsilon) - Z(\theta_j - \epsilon)}{2\epsilon}
    + \mathcal{O}(\epsilon^2),
    \label{eq:1st_order_finite_difference}
\end{align}
where we have used the shorhand notation $Z(\boldsymbol{\theta}, t) = Z(\theta_1, \theta_2, \ldots \theta_j, \ldots  \theta_M,t) = Z(\theta_j)$. Note, the error of the approximation in Eq.~(\ref{eq:1st_order_finite_difference}) can be made negligibly small by a sufficiently small selection of $\epsilon$, but not too small to induce round-off errors. In a similar manner, the second order derivative reads

\begin{align}
    \frac{\partial^2 Z(\theta_j)}{\partial \theta_j^2}
    = \frac{Z(\theta_j + \epsilon) - 2Z(\theta_j) + Z(\theta_j - \epsilon)}{\epsilon^2}
    + \mathcal{O}(\epsilon^2).
    \label{eq:2nd_order_finite_difference}
\end{align}

Note that Eq.~(\ref{eq:1st_order_finite_difference}) and (\ref{eq:2nd_order_finite_difference}) requires $1+2M$ function evaluations, where $M$ again denotes the number of system-parameters. That is, for applications where we do not model covariances (i.e. assume independent parameters), and only need the mean value, our method scales linearly with the number of function evaluations. 

If we also needed the mixed derivatives, a finite difference scheme is given by

\begin{align}\nonumber
    &\frac{\partial^2 Z(\theta_j, \theta_i)}{\partial \theta_j \partial \theta_i}
    = \frac{1}{\epsilon^2} 
    \Big[
    Z(\theta_j + \epsilon, \theta_i + \epsilon)
    -Z(\theta_j + \epsilon, \theta_i)
    -Z(\theta_j, \theta_i + \epsilon)\\
    &+2Z(\theta_j, \theta_i)
    -Z(\theta_j - \epsilon, \theta_i)
    -Z(\theta_j, \theta_i - \epsilon)
    +Z(\theta_j - \epsilon, \theta_i - \epsilon) 
    \Big]
    +\mathcal{O}(\epsilon^2).
    \label{eq:2nd_order_finite_difference_mixed}
\end{align}
Note, this only requires two additional function evaluations for each of the $\frac{1}{2}M(M-1)$ possible combinations. This means that the overall number of evaluations becomes $1+M+M^2$. Note that although finite difference schemes are often easier to implement, they are not as numerically stable as calculating the exact derivatives, which does not depend on a choice of $\epsilon$. Note in the main text, Section~\ref{sec:scalling}, we compare calculating the mean with independent parameters, and so we use a scaling of $1+2M$. 

\subsection*{Experimental details}

The details of the experiment are as follows: First, we initialize the qubit in the $\ket{0}$ state. We then drive the qubit on resonance with a series of Gaussian pulses of constant width (the pulse is truncated at $\pm6\sigma$, where $\sigma = 75$~ns) and varying amplitude, which effectively realizes the two level Hamiltonian in the main text with $\Delta = 0$ and varying $\Omega_\mathrm{max}$. The qubit is then measured and a discriminator is applied such that we obtain a zero or one depending on if the qubit was in the ground or excited state, respectively. This experiment, and all other experimental runs reported on in this manuscript, is repeated $1024$ times to obtain statistics, and the resulting curve is fit to a sinusoid and scaled between $\braket{z} = \pm 1$, which allows us to maintain consistency with theory in the face of experimental systematics that result in finite contrast. From this, we can determine the pulse amplitudes $A_\Theta$ required to obtain a given $X$ rotation $\Theta$. This allows us to calibrate the system.

The Gaussian pulses found above can also be used to form composite BB1 pulse sequences~\cite{wimperis1994broadband}, which take the form
\begin{equation}
    \label{eq:BB1}
    R_\mathrm{BB1}(\Theta) = R_{\pi, \phi(\Theta)} R_{2\pi, 3\phi(\Theta)} R_{\pi, \phi(\Theta)} R_{\theta, 0}
\end{equation}
where $R_{\theta, \phi}$ defines a Gaussian pulse with a phase $\phi$ that results in a rotation $\theta$. The optimal phase $\phi$ for a desired final rotation $\Theta$ is given by $\phi(\Theta) = \cos^{-1}{(-\Theta/4\pi)}$ \cite{wimperis1994broadband}. 


\twocolumngrid

\bibliography{refs}

\end{document}